\documentclass{article}

\usepackage{cite}
\begin{document}
\begin{center}
\Large{\textbf{A Gauss Elimination  Method for Resonances.}}\par
\end{center}
\vspace{.75cm}
\Large
\hspace{1cm}
\begin{center}
\large{John P.Killingbeck$^{*}$, Alain Grosjean.}\par
\end{center}
\large
\noindent
\begin{center}
\textsl{Institut Utinam (CNRS,UMR 6213)
, 41 bis Avenue de l'Observatoire, B.P.1615, 25010 Besan\c{c}on Cedex, France}\par
\end{center}
\begin{center}
$^{*}$\textsl{Mathematics Centre, University of Hull, Hull HU6 7RX,U.K}
\end{center}
\vspace{2cm}
\large
\vspace{1cm}

\Large
\underline{ABSTRACT}\\
\large
A Gaussian elimination form of inverse iteration within the complex coordinate approach
is shown to produce a simple uniform method of finding both real bound state energies
and complex resonant state energies  for
several problems which have been treated by a variety of methods in the literature. The 
energy shift method for expectation values is shown to be a useful diagnostic tool.

\section{INTRODUCTION}
 
Several recent works have shown that the resonant state energies for some
systems can be calculated by using a straightforward approach in which the
equations appearing in a traditional bound state calculation are modified
by the simple procedure of making an imbedded parameter become a complex variable. Thus, for
example, the coefficient $\beta$ in a wavefunction factor of type e$^{-\beta\frac{x^{2}}{2}}$
for perturbed oscillator problems or the coefficient Z in a wavefunction factor of type e$^{-Zr}$
for perturbed hydrogen atom problems are usually varied to optimize the calculation of bound state
energies; however they also make it possible to find complex resonant state energies when they are 
given complex values.
This simple complexification approach has been shown to work for hypervirial perturbation theory
(HVPT)\cite{kill1}, for matrix diagonalization methods \cite{kill2} and for the 
Hill-series method \cite{kill3}. 
As a remarkable example of this approach it has been found that for a
perturbed hydrogen atom a simple moment method can give either real
Zeeman effect energies ( with Z real ) or complex Stark effect energies ( with Z 
complex ) \cite{kill4}.\\
The traditional complex rotation method for locating resonant state energies involves 
a rotation of all the operators in the Hamiltonian ( including the kinetic energy operator ) 
and leads to a complex symmetric matrix eigenvalue problem. There has been some debate in the previous 
literature about the situations in which the complex rotation method is equivalent to the more
simple complex basis approach\cite{baum,gya}, with the authors of \cite{gya} claiming that the complex
basis approach is the more fundamental one. However, for analytic potentials it is generally agreed 
that the matrices for the two approaches are related by a similarity transformation and so lead to
identical spectra. As explained in section 2 the complex basis approach is more easy to use and
in the present work we use it to look at the problem of finding only a selected few of the eigenvalues of
a complex symmetric matrix.\\
In \cite{kill2} the complex symmetric matrix was tranformed by means of a sequence of  
simple similarity transformations which is based on
first order perturbation theory and which is similar to the complex Jacobi approach. The method is
sufficiently general to work even for non-symmetric matrices but has the disadvantages that it 
sets out to find the full spectrum and that it fills up the matrix as it proceeds.
For many test problems we only require a few low-lying eigenvalues  and are dealing with an initial matrix
which has only a small bandwidth.
As a method for solving systems of linear equations the traditional 
Gaussian elimination method has the special feature that it preserves both the bandwidth and
the symmetry of the initial square matrix during the reduction process.
The appropriate way to exploit this property while finding only a few eigenvalues is to extract 
the eigenvalues and eigencolumns by using a complex version of the Gaussian elimination method to
perform inverse iteration on the complex symmetric matrix. By finding $(H-E)^{-N}Y$ for some starting
column and a sufficiently large N we can extract the eigenvalue closest to E.
Because the matrix bandwidth and symmetry is undisturbed during the process the matrix can be stored
in a very compact form. Such an approach is more simple than the use of a complex version of other 
methods such as the Lanczos method, which, although it finally reduces the matrix to tridiagonal form, 
would require a filter diagonalization approach and the conversion of the basis set by the operator
$(H-E)^{-1}$ to perform the calculations reported here \cite{ala1}. For small bandwidths it is obviously
easier to deal with the matrix directly, as we do here.\\
In the present work we use our method to find the complex energies of resonant states of real potentials as 
well as the real energies for PT symmetric 
potentials. Section 2 describes the method and later sections give
results for several systems, some of which have been treated 
in the previous literature. Section 3 produces both bound and resonant state energies for
a triple well system which was recently treated by a Hill-series method.
Section 4 treats a PT invariant Hamiltonian which
has been much analyzed in the literature and for which broken symmetry effects lead to the presence 
of both real and complex eigenvalues in the spectrum. Section 5 deals with a cubically perturbed oscillator
which has previously been used as an example of hyperasymptotic analysis.
Section 6 shows that our approach can handle some unorthodox types of resonance.
Section 7 deals with only bound states but demonstrates that it is not necessary to use basis functions of definite 
parity to treat a centrosymmetric potential. The numerical results show an initial quasi-convergence
of the eigenvalues to the energy given by HVTP. Section 8 gives a brief discussion.

\section{THE METHOD OF CALCULATION.}

All the systems treated in this work can be handled in a matrix approach by using
an harmonic oscillator basis set. The advantage of using the complex basis approach is that it can be implemented
by using a very simple prescription; we write down the real energy bound state theory and replace the relevant real 
parameter throughout by a complex one. Thus, if we use a set of basis functions which are the eigenfunctions of
the reference Hamiltonian

\begin{equation} 
        H(W)=-\alpha D^{2}+Wx^{2}                  
\end{equation}
 
then we know that the energy of state n is  $E_{n}=(2n+1)(W\alpha)^{\frac{1}{2}}$ and that the coordinate 
operator $x$ has the matrix element

\begin{equation} 
        <n|x|n+1>=(\frac{\alpha}{4W})^{\frac{1}{4}}(n+1)^{\frac{1}{2}}                
\end{equation}
 
To handle complex eigenvalue problems we simply make W complex in the formulae given above.
This will require the extraction of complex square and fourth roots to evaluate the constants appearing in the formulae.
The matrices for
higher powers of $x$ can be constructed by using the complex form of the algebraic formula for the
appropriate matrix element or by direct numerical matrix multiplication. To find the exact matrix elements
by matrix multiplication we proceed by avoiding the kind of edge effects which would give only approximations
similar to those of the DVR or HEG approach \cite{kill2}.
Thus, for example, to find the $x^{3}$ matrix of N$\times$N type we would form 
the $x$ matrix of (N+3)$\times$(N+3) type and then reduce by 1 the dimension
at each step of the multiplications. Alternatively, the final matrix can be computed row
by row, using a long row which is filled up with the elements of the desired multiple product, 
which are then loaded into the matrix.
We can illustrate the use of the eigenfunctions  of (1) by showing the partitioning of a typical
perturbed oscillator Hamiltonian ( for a cubic perturbation ):

\begin{equation} 
        H=-\alpha D^{2}+Ux^{2}+Vx^{3}=[-\alpha D^{2}+Wx^{2}]+(U-W)x^{2}+Vx^{3}   
\end{equation}
 
The term in square brackets is diagonal with diagonal elements E$_{n}$, while the matrix elements of $x^{2}$
and $x^{3}$ are found as explained above. For brevity (3) shows the real variable case. For a resonance W becomes
complex (eg W=WR+iWI), while V can be real or imaginary depending on the problem studied. Most of the oscillator resonant states
treated in the literature issue from states which are initially oscillator bound states before the action of
some perturbing potential. For such states the WR is usually kept equal to the positive real value of W which is associated
with the initial bound state, with WI being varied; when EI is small this choice suffices to give accurate results.\\
To search for a complex eigenvalue in the neighbourhood of energy E we use Gaussian elimination to find a
sequence of columns $X(n)$ which obey the equation

\begin{equation} 
           X(n+1)=(H-E)^{-1}X(n)
\end{equation}
 
The calculation proceeds as follows. At each stage we set Y=X(n) and then solve the linear equation systeme
$(H-E)X=Y$ for X using Gaussian elimination. We then set X(n+1)=X. In all the calculations reported here it sufficed 
to use simple Gauss elimination without pivoting and with a real E, since the complex arithmetic quickly introduced
the imaginary part EI for the sought eigenvalue, which is not very large for the lower eigenvalues. The initial
column X(0) is a column with every element given the real value 1. At
each iteration the obtained column X(n+1) is scaled to make its first element equal to the real 
value 1 and then the current eigenvalue estimate is found by evaluating the first row
of the complex product HX(n+1). Since complex arithmetic is being used it is possible to seek faster convergence
by using the latest complex E in the E position in (4); 
however  this loses the advantage of having a fixed 
form of the reduced matrix $(H-E)$ throughout the calculation. After N steps of the process we have obtained
$(H-E)^{-N}X(0)$, which is dominated by the contribution from the eigenvector with its eigenvalue closest to the E
value being used. To find several of the low 
eigenvalues E is gradually increased to scan the relevant energy range. The are three ways to check the stability 
of the results. First, the matrix dimension can be gradually increased to check for a stable convergence to a limit,
checking that the limit has a negligible dependence on W. Second, as E is varied in small steps DE through the scanned
region the same eigenvalue should emerge over several successive steps of E if the calculation is initialized at each
step. Third, the reference row for the evaluation of E can be varied to see whether the eigenvalue varies. (The essential 
point is that the basis function associated with the reference row must make a reasonably large contribution to the
eigencolumn).

\section{AN INTERESTING TRIPLE WELL SYSTEM.}

A recent study of the complexified form of the Hill-series technique \cite{kill3}
included some calculations for a special triple well Hamiltonian

\begin{equation} 
        H(g)=-D^{2}+x^{2}-2g^{2}x^{4}+g^{4}x^{6}   
\end{equation}
 
which was introduced in \cite{ben} and later used to illustrate how the use 
of complex coordinates in a matrix diagonalization approach can describe tunnelling effects
in bound systems \cite{mois}. The Ricatti-Pad\'{e} 
method can also describe these tunnelling effects \cite{fern}.
By varying the imbedded
parameter $\beta$ in the factor e$^{-\beta x^{2}/2}$ appearing in the Hill-series formalism
it was found possible to produce both
real and complex eigenvalues for H(g) at small g values.
From a physical point of view the resonances can be regarded as referring to
the outwards tunnelling of a wavepacket initially sited in the inner well;
from a mathematical point of view they are associated with a complex scaled 
Hamiltonian \cite{ben}. The real eigenvalues are just the energies of the traditional bound states
which would be expected for a potential which rises towards infinity 
at large $x$ values. The system described by (5) thus provides an obvious test for the method of
the present work ( and the prescription described at the start of section 2 ). The matrix of H(g)
requires the first three powers of the $x^{2}$ matrix for its construction. The matrix
elements H(J,K) with K=J to J+3 are the only ones needing storage, since the matrix is symmetric  
and we chose a basis with a definite even or odd parity. With this choice the compact storage scheme uses a linear
array HC for wich the H(J,K) element is stored at the element HC(M) such that

\begin{equation} 
        H(J,K)=HC(3J+K-3) \hspace{1cm}      (K=J\hspace{0.1cm} to\hspace{0.1cm} J+3)  
\end{equation}
 
This coding has, of course, to be used in the various matrix operations and a copy of the matrix
is used in the elimination process ( which destroys the original matrix ). The use of 
compact storage would make it possible to use very large basis sets for a symmetric matrix with
a small bandwidth. Although we only use dimensions up to about 200 in our examples, the approach avoids 
the large number of arithmetic operations involved in our previous method \cite{kill2} and so is less
subject to the effects of rounding errors when ordinary double precision is used. The full power of the
compact matrix operations is obviously not needed for the case of the smooth potentials treated here but is
necessary for handling singular perturbations of the oscillator, where it makes basis sizes of up to 20,000
attainable in a simple procedure ( work in preparation ).
Numerical calculations showed that the complex resonance
energies E=ER+EI with ER close to 1, as given in Table 3 of \cite{kill3}, were given to double
precision by using the W value (1,15). The bound state energies of Table 4 of 
\cite{kill3} were found by using W=(1,0). For these two cases 
an even parity basis set was used.
We continued the calculation to give extra resonances which were not studied 
in previous works. The lowest four resonances found are shown in Table 1, 
for several g values. The barrier between the inner and outer wells has a height equal to 
$\frac{4}{27}g^{2}$; the value of EI increases markedly as ER passes through that value.
A calculation using complex HVPT \cite{kill1}, although less accurate, agreed well
with the matrix calculations. This auxiliary perturbation calculation  
can identify the particular state from
which a resonance arises and so indicate the region of E to scan in the matrix calculation.
Table 2 shows some real bound state energies found for several g values
by setting (WR,WI)=(1,0). The states with an energy just below 2 were found by
noting that at the outer minimum the potential has a leading term $4x^{2}$. It thus seems likely that they 
are associated with localized states in the outer well. That this is so was confirmed in two different ways.
First, an HVPT calculation with its origin at the centre of the outer well ( and using the Taylor expansion
about that point ) gave energies which agree closely with those obtained from the matrix calculation.
Second, within the matrix calculation itself the small perturbing term 0.00005$x^{2}$ 
was in turn added to and subtracted from the potential; the perturbed eigenvalue was calculated for each
case. The results then gave an accurate estimate of the expectation value $<x^{2}>=\frac{dE}{dV_{2}}$
and the result indicated that the wavefunction is indeed concentrated near the centre of the outer well.
This energy shift approach to finding expectation values is widely applicable and here is an efficient alternative
to the traditional method of working out $<x^{2}>$ by using the matrix of $x^{2}$ together with
the eigencolumn, particularly since the eigencolumn would have to be normalized in a preliminary step.

\section{A SYSTEM WITH PT INVARIANCE.}

The Schroedinger equation

\begin{equation} 
        (-D^{2}+Aix^{3}+Bix)\Psi=(ER+EI)\Psi   
\end{equation}
 
has been studied by several authors [ e.g \cite{han,bran}]. The PT symmetry of the operator in (7)
suggests that the eigenvalues can be real ( with EI=0 ). When B=0 the spectrum is indeed
entirely real but for non-zero B it is possible to have complex eigenvalues, as demonstrated numerically 
in the moment method calculations of \cite{han}. It is thus a suitable test for the method of this work to
see whether it can describe both the real eigenvalues and the complex symmetry breaking eigenvalues for the
operator in (7), by analogy with the way in which it gave both real and complex eigenvalues for the problem
treated in section 3. To set up the matrix we require the first three powers of the matrix of $x$.
The $x^{2}$ matrix arises because we need to adopt what is essentially a renormalizing approach, in which a term
-(WR+iWI)$x^{2}$ must be included in the perturbing potential in order to cancel the term which is implicit in
the operator (2) associated with the basis set.
Table 3 shows some typical results for the problem, which are more accurate than those published in previous works.
The results for the special case A=1, B=-3, -4, -5 were obtained by 
gradually increasing E in the operator (H - E) of the Gaussian elimination process.  
If a complex W is used in an attempt to find a real eigenvalue then the EI value obtained decreases as the iterations
proceed, finally oscillating with an amplitude of roughly $10^{-14}$ at the standard level of double precision which
we used. The method of calculating expectation values explained in Section 3 can be used for this problem.
For example, if we set A=-1 and B=0 ( to get a real spectrum )
then we can add a very small $x^{2}$ perturbing term to the potential to show that 
for the potential $\beta x^{2}-ix^{3}$ the lowest order correction
term for the ground state energy is $1.9669085\beta ^{2}$.

\section{THE CUBICALLY PERTURBED OSCILLATOR.}

Alvarez and Casares \cite{alv} described the use of hyperasymptotics to find the complex 
eigenvalues for the operator

\begin{equation} 
        H(g,\phi)=-\frac{1}{2}(D^{2}+x^{2})+gx^{3}exp(i\phi)   
\end{equation}
 
Using a complex perturbation parameter in the traditional Rayleigh-Schroedinger perturbation series for the 
energy would obviously give $E(g,-\phi)$ as the complex conjugate of $E(g,\phi)$. However the authors of \cite{alv}
pointed out that the presence of a Stokes line at $\phi$=0 means that the hyperasymptotic correction to the perturbative
result is needed for negative $\phi$ values, so that the complex conjugation symmetry is destroyed.
It was later
shown that the complex HVPT can give the corrected results directly for some distance into
the negative $\phi$ region \cite{kill5}. 
Table 4 shows results obtained by the Gaussian elimination method at g=0.1. 
Although, as explained in the introduction, the complex coordinate method is equivalent to ( though more simple than )
the complex scaling matrix method used to obtain reference values in \cite{alv}, our results are more accurate than those 
given in \cite{alv} and cover a finer grid of values of the angle $\phi$.

\section{SOME UNORTHODOX RESONANCES.}

In \cite{gom} some complex energies were calculated for the Hamiltonian

\begin{equation} 
        H(A,B)=-D^{2}+x^{M}-\lambda x^{N}   
\end{equation}
 
The most commonly studied case in the literature has M=2, N=4 but in \cite{gom} the case M=4, N=6 was also
discussed. Table 5 shows some results for two sets of (M,N) values, as obtained by the Gaussian elimination method.
For negative $\lambda$ values, of course, bound states exist and the use of a real W value in the basis set can produce these.

\section{SYMMETRY AND THE DOUBLE WELL PROBLEM.}

All the preceding examples studied in this work have involved complex resonances but our final example deals with
the ordinary bound states associated with the double well Hamiltonian

\begin{equation} 
        H(\lambda)=-D^{2}-x^{2}+\frac{1}{2}\lambda ^{2}x^{4}   
\end{equation}
 
When $\lambda$ is small this Hamiltonian will have close pairs of states of opposite parity at the bottom of its spectrum.
The dominant tradition in matrix approaches is to use the symmetry centre $x$=0 as origin and so to build the even or odd 
parity directly into the basis functions used to set up the matrix of the Hamiltonian. Thus, for example, by using an 
oscillator basis with W=(1,0) in separate even or odd parity calculations based on the $x$=0 origin our method gives
the low lying pairs of levels. Some authors keep the origin at $x$=0 but use a basis which consists of even or odd combinations of 
oscillator functions centred at the well minima at [$x=\pm\frac{1}{\lambda}$] \cite{arias}.
While such basis functions show a strong resemblance to the actual eigenfunctions they have the disadvantage of being
non-orthogonal, which complicates the matrix eigenvalue problem. We tried a calculation in which only the oscillator functions
centred on $x=\frac{1}{\lambda}$ were used, to see whether the potential itself would introduce the centrosymmetry into
the basis and generate the correct partner peak on the other side of the origin. To perform this calculation with the point
$x=\frac{1}{\lambda}$ as origin we need to use the correct potential, which has admixtures of both even and odd parity. It
is found by performing the Taylor expansion at $x=\frac{1}{\lambda}$ of the potential appearing in (10) and takes the form

\begin{equation} 
        V=-\frac{1}{2\lambda ^{2}}+2x^{2}+2\lambda x^{3}+\frac{1}{2}\lambda^{2}x^{4}   
\end{equation}
 
This potential naturally suggests the choice W=(2,0) for the basis functions, with both even and odd functions
being included in the basis. Table 6 shows the results obtained; as the dimension is increased the energies tend 
correctly to those for the lowest even and odd pair of levels.
This effect is explainable in terms of the specially favourable properties of harmonic oscillator basis 
functions. The virial theorem shows that the value of $<x^{2}>$ for an oscillator function is proportional to its energy
i.e. is proportional to 2n+1. Thus, as more functions are added to the basis set we reach a stage at which the basis functions 
can have an appreciable overlap with the region in the opposite well and so make it possible to describe the partner peak in 
the wavefunction. This physical argument describes quite well what happens in the numerical calculation.
To establish that the correct lowest eigenvalue is indeed associated with the "double peak" symmetric wavefunction we can
add small $x$ and $x^{2}$ terms to the potential, as explained previously, to find $<x>$ and $<x^{2}>$ and so reveal the presence
of the second peak at a distance of about $\frac{2}{\lambda}$ from our origin at the point $x=\frac{1}{\lambda}$
An interesting extra effect appeared in the numerical results.
For the deeper double well ( with $\lambda=0.3$ ) the results of table 6 show that the single eigenvalue near -4.2 reaches 
a semi-converged value of roughly -4.190234 as the matrix dimension increases. This energy is the average of the correct
lowest even and odd state energies
and is also the energy obtained when a hypervirial perturbation calculation is performed using the potential in (11).
As the matrix dimension is further increased the original single level eventually begins to descend ( to become the lowest level )
and is joined by a level 
which moves down from above to form the upper level of the lowest lying doublet. Many years ago Seznec and Zinn-Justin \cite{sezn}
suggested that perturbation theory based on the well centre should give an energy close to the average of the even
and odd state energies, and HVPT does this. It would be an interesting calculation  ( albeit requiring very high precision ) to
see what trajectory would be followed by the lowest eigenvalue given by a matrix-based perturbation approach which uses exactly
the same matrix elements as those of the matrix diagonalization approach based on $x=\frac{1}{\lambda}$. In particular
one could ask whether at a sufficiently high order the perturbation approach would "break through" the region of semi-convergence and 
descend with the true matrix eigenvalue so as to arrive at the correct even parity ground state energy.

\section{CONCLUSION.}

The results for the specimen systems studied in this work show that by combining Gaussian elimination with
inverse iteration using complex basis functions we can make an accurate study of either real or complex eigenvalues 
in a desired region of the spectrum while retaining the small bandwidth of the Hamiltonian matrix which is 
characteristic of many of the systems which have been studied by various different methods in the literature \cite{present}. For
systems which can have broken PT symmetry the ability to handle both real and complex energies by simple adjusting the reference 
parameter W is particularly useful. The relationship between a perturbation approach and the unexpected semi-convergence of the 
matrix eigenvalues for the double well problem which we found in section 7 is clearly something which merits further investigation.

\newpage

\underline{TABLE 1.}
The lowest three resonance energies for the Hamiltonian H(g) of
equation (6), obtained using W=(1,15) and a matrix dimension of 150. The parity P
is either even (E) or odd (O).\\

\vspace{0.5cm}

\begin{tabular}{|  r@{.}l |c|  r@{.}l|          r@{.}l|}
\hline

\multicolumn{2}{|c|}{g}
&\multicolumn{1}{c|}{P}
&\multicolumn{2}{c|}{ER}
&\multicolumn{2}{c|}{EI}\\
\hline

0&20&E&0&9325571582478&7&94775543926(-5)\\
&&O&2&6156743444473&1&21030060549(-2)\\
&&E&3&8713869659323&1&99483314620(-1)\\
0&24&E&0&8944205532099&2&42463284005(-3)\\
&&O&2&3894780354803&1&11999490115(-1)\\
&&E&3&4581087741326&6&60180783826(-1)\\
0&28&E&0&8433344239234&1&59158594653(-2)\\
&&O&2&1950967814330&3&02661677759(-1)\\
&&E&3&2043949873518&1&18854425437(0)\\

\hline

\end{tabular}

\newpage

\underline{TABLE 2.}
The lowest four bound state energies for the Hamiltonian H(g)
of equation (6), obtained using W=(1,0) and matrix dimension 150. The parity
labels are as in table 1. The $<x^{2}>$ values are found by the energy shift method.\\

\vspace{0.5cm}

\begin{tabular}{|  r@{.}l |c|  r@{.}l|          r@{.}l|}
\hline

\multicolumn{2}{|c|}{g}
&\multicolumn{1}{c|}{P}
&\multicolumn{2}{c|}{E}
&\multicolumn{2}{c|}{$<x^{2}>$}\\
\hline

0&20&E&0&93247629196422&0&596\\
&&O&1&81996584353442&22&315\\
&&E&1&82258016776947&22&423\\
&&O&2&62828330994496&2&438\\
0&24&E&0&89204244181975&0&768\\
&&O&1&69073242323339&13&508\\
&&E&1&73636556408804&14&348\\
&&O&2&53097937792111&4&093\\
0&28&E&0&82917630720481&1&121\\
&&O&1&53456526498005&8&495\\
&&E&1&70854344684062&9&587\\
&&O&2&64073480349469&4&817\\

\hline

\end{tabular}

\newpage

\underline{TABLE 3.}
Low lying eigenvalues for the operator $-D^{2}+Aix^{3}+Bix$,
calculated using a W value of(1,15) and a matrix dimension of 150 with an even parity
basis. Three broken symmetry states with complex energy are shown.\\

\vspace{0.5cm}

\begin{tabular}{|c|c|  r@{.}l |  r@{.}l|}  
\hline

\multicolumn{1}{|c|}{A}
&\multicolumn{1}{c|}{B}
&\multicolumn{2}{c|}{ER}
&\multicolumn{2}{c|}{EI}\\
\hline

1&0&1&15626707198811&&\\
1&0&4&10922875280966&&\\
1&0&7&5622738549787&&\\
1&0&11&3144218201962&&\\
1&0&15&291553750390&&\\
1&-5&1&34334319874918&-2&9073906160965\\
&&3&43138320167211&&\\
&&5&16788868578734&&\\
1&-4&1&24865673359469&-1&7617193016512\\
&&3&50876560739555&&\\
&&6&37980520633110&&\\
1&-3&1&22584757671327&-0&76002247143487\\
&&4&33343983644352&&\\
&&7&52519195567867&&\\

\hline

\end{tabular}

\newpage

\underline{TABLE 4.}
Groundstate complex eigenvalues for the cubic perturbed
oscillator with the Hamiltonian H(g,$\phi$) given in equation (8), for g=0.1.
An even parity basis with W set equal to (0.5,0.5)was used, with a matrix dimension of 150\\ 

\vspace{0.5cm}

\begin{tabular}{|r@{.}l |  r@{.}l |  r@{.}l|}  
\hline

\multicolumn{2}{|c|}{$\varphi$}
&\multicolumn{2}{c|}{ER}
&\multicolumn{2}{c|}{EI}\\
\hline

-0&10&0&4848327348572&3&621442463000(-3)\\
-0&08&0&4846443760119&2&91525529968(-3)\\
-0&06&0&4844977122642&2&19506948166(-3)\\
-0&04&0&4843938809947&1&46508620844(-3)\\
-0&02&0&4843333450837&7&29406327924(-4)\\
0&00&0&4843159970041&-8&06020950000(-6)\\
0&02&0&4843412576766&-7&43653067984(-4)\\
0&04&0&4844081666578&-1&47399596288(-3)\\
0&06&0&4845154620899&-2&19601304015(-3)\\
0&08&0&4846616500444&-2&90693218571(-3)\\
0&10&0&4848450636272&-3&60427916939(-3)\\

\hline

\end{tabular}

\newpage

\underline{TABLE 5.}
The lowest even and odd parity resonances for operator
$-D^{2}+x^{M}-\lambda x^{N}$,calculated using W=(1,1) and matrix dimension 150.
The even parity energy is given first\\
\vspace{0.5cm}

\begin{tabular}{|c|c|r@{.}l |  r@{.}l |  r@{.}l|}  
\hline

\multicolumn{1}{|c|}{M}
&\multicolumn{1}{c|}{N}
&\multicolumn{2}{c|}{}
&\multicolumn{2}{c|}{ER}
&\multicolumn{2}{c|}{EI}\\
\hline

2&6&0&02&0&9520462653053309&0&01402573778245021\\
2&6&0&02&2&712788208122200&0&2013898488709008\\
2&6&0&04&0&9193107387010802&0&05273643153667654\\
2&6&0&04&2&654858021276504&0&4633716332349316\\
2&6&0&06&0&9033613239572396&0&09127664934058118\\
2&6&0&06&2&660081776872349&0&6557404144178417\\
2&6&0&08&0&8958197460011326&0&1254636428092339\\
2&6&0&08&2&682850763658098&0&8064038265203445\\
2&6&0&10&0&8927457964926816&0&1555432433598369\\
2&6&0&10&2&711585309963890&0&9306087023690415\\
4&6&0&00&1&060362090484183&0&0\\
4&6&0&00&3&799673029801394&0&0\\
4&6&0&04&1&038002353717577&0&0\\
4&6&0&04&3&699156060168530&0&0\\
4&6&0&08&1&012731445721011&6&25221492542814(-8)\\
4&6&0&08&3&581700216602671&1&288075605381661(-6)\\
4&6&0&12&0&9826725857365955&1&594662368407940(-4)\\
4&6&0&12&3&431460367418278&2&739336833818093(-3)\\
4&6&0&16&0&9440969584873676&3&992421290169972(-3)\\
4&6&0&16&3&22859327560889&5&18915247310509(-2)\\
4&6&0&20&0&8995394462905228&2&019133790314381(-2)\\
4&6&0&20&3&036336773026575&1&920564090658734(-1)\\

\hline

\end{tabular}

\newpage

\underline{TABLE 6.}
Energy levels for the double well Hamiltonian using the potential of equation (11), with 
origin at $x=\frac{1}{\lambda}$, for the cases $\lambda$=0.3 and $\lambda=0.4$, as a function of
the matrix dimension ND. 
The E scanning region used (-4.3,-4.1) for $\lambda$=0.3 
(-1.9,-1.7) for $\lambda$=0.4,  in steps DE of 0.02.\\

\vspace{0.5cm}

\begin{tabular}{|c|  r@{.}l |  r@{.}l|}  
\hline

\multicolumn{1}{|c|}{ND}
&\multicolumn{2}{c|}{E(A)}
&\multicolumn{2}{c|}{E(B)}\\
\hline

10&-4&1902095978175&-1&8054955213226\\
20&-4&1902336009106&-1&8068973696846\\
30&-4&1902339345051&-1&7751402016719\\
&&&-1&8207465095929\\
40&-4&1902342389732&-1&7847047589878\\
&&&-1&8267498723262\\
50&-4&1902508125434&-1&7847050286351\\
&&&-1&8267501124629\\
60&-4&1899127461966&-1&7847050286292\\
&-4&1905545952753&-1&8267501124656\\
70&-4&1899128809639&-1&7847050286292\\
&-4&1905545952753&-1&8267501124656\\
80&-4&1899128809648&-1&7847050186292\\
&-4&1905545952761&-1&8267501124656\\

\hline

\end{tabular}

\newpage

\end{document}